\documentclass[letterpaper, 10 pt, conference]{ieeeconf}  
\IEEEoverridecommandlockouts
\makeatletter
\let\NAT@parse\undefined
\makeatother
\usepackage{amsmath}
\usepackage{amssymb,bm}
\usepackage{amsfonts}
\usepackage{enumerate}
\usepackage{tabulary}
\usepackage{multirow}
\usepackage{cite}
\usepackage{lipsum} 
\usepackage{verbatim}
\usepackage[hidelinks]{hyperref}
\usepackage{url}
\usepackage{balance}
\usepackage{graphicx}
\usepackage{subcaption}
\usepackage[font=footnotesize]{caption}
\usepackage{epstopdf}
\usepackage[misc]{ifsym}
\usepackage{color}
\usepackage{xcolor}
\usepackage{xxcolor}
\usepackage{tikz}
\usepackage{pgf}
\usepackage{pgfplots}
\usepackage{float}
\usepackage{siunitx}
\usepackage{ifthen}
\usepackage{forloop}
\usepackage{listings}
\usepackage{booktabs}
\usepackage[lined,boxed,commentsnumbered,linesnumbered]{algorithm2e}

\usepackage[utf8]{inputenc}
\usepackage{pgfplots}
\DeclareUnicodeCharacter{2212}{−}
\usepgfplotslibrary{groupplots,dateplot}
\usetikzlibrary{patterns,shapes.arrows}
\pgfplotsset{compat=newest}

\hypersetup{
  colorlinks    = true, % Colours links instead of ugly boxes
  urlcolor      = blue, % Colour for external hyperlinks
  linkcolor     = blue, % Colour of internal links
  citecolor     = blue   % Colour of citations
}
\definecolor{codegreen}{rgb}{0,0.6,0}
\definecolor{codepurple}{rgb}{0.58,0,0.82}
\definecolor{backcolour}{rgb}{0.95,0.95,0.92}
\lstdefinestyle{buzz}{
    backgroundcolor=\color{black!5},   
    commentstyle=\color{codegreen},
    keywordstyle=\color{blue},
    numberstyle=\tiny\color{black!30},
    stringstyle=\color{codepurple},
    basicstyle=\footnotesize\ttfamily,
    breakatwhitespace=false,         
    breaklines=true,                 
    captionpos=b,                    
    keepspaces=true,                 
    numbers=left,                    
    numbersep=5pt,                  
    showspaces=false,                
    showstringspaces=false,
    showtabs=false,                  
    tabsize=2,
}
\renewcommand{\vec}[1]{#1}
\newcommand{\xdot}{\dot{x}}
\newcommand{\x}{x}
\renewcommand{\u}{u}
\newcommand{\f}{f}
\newcommand{\g}{g}

\renewcommand{\u}{\vec{u}}

\newcommand{\set}[1]{\mathbb{#1}}
\newcommand{\R}{\mathbb{R}}
\newcommand{\N}{\mathbb{N}}
\newcommand{\K}{\mathcal{K}}

\usepackage{stackengine}

\newtheorem{definition}{Definition}

\newcommand{\todo}[1]{\textcolor{black}{#1}}

\SetAlCapSkip{-0.25em}
\title{\LARGE \bf
Practical Considerations for Discrete-Time Implementations of Continuous-Time Control Barrier Function-Based Safety Filters
}

\author{Lukas Brunke, Siqi Zhou, Mingxuan Che, and Angela P. Schoellig
\thanks{The authors are with the 
\href{http://www.learnsyslab.org}{Learning Systems and Robotics Lab} at the Technical University of Munich, Germany and the 
	University of Toronto, Canada. The authors are also affiliated with
	the Munich Institute of Robotics and Machine Intelligence~(MIRMI), the  University of Toronto Robotics Institute, and the Vector Institute for Artificial Intelligence. Emails:
	\{lukas.brunke, siqi.zhou, mingxuan.che, angela.schoellig\}@tum.de}%
}

\begin{document}
\maketitle
\thispagestyle{empty}
\pagestyle{empty}

%%%%%%%%%%%%%%%%%%%%%%%%%%%%%%%%%%%%%%%%%%%%%%%%

\begin{abstract}
  %~150 words
Safety filters based on control barrier functions~(CBFs) have become a popular method to guarantee safety for uncertified control policies, e.g., as resulting from reinforcement learning. 
Here, safety is defined as staying in a pre-defined set, the safe set, that adheres to the system's state constraints, e.g., as given by lane boundaries for a self-driving vehicle.
In this paper, we examine one commonly overlooked problem that arises in practical implementations of continuous-time CBF-based safety filters.
In particular, we look at the issues caused by discrete-time implementations of the continuous-time CBF-based safety filter, especially for cases where the magnitude of the Lie derivative of the CBF with respect to the control input is zero or close to zero. 
When overlooked, this filter can result in undesirable chattering effects or constraint violations. 
In this work, we propose three mitigation strategies that allow us to use a continuous-time safety filter in a discrete-time implementation with a local relative degree.
Using these strategies in augmented CBF-based safety filters, we achieve safety for all states in the safe set by either using an additional penalty term in the safety filtering objective or modifying the CBF such that those undesired states are not encountered during closed-loop operation. 
We demonstrate the presented issue and validate our three proposed mitigation strategies in simulation and on a real-world quadrotor. 
\end{abstract}

%%%%%%%%%%%%%%%%%%%%%%%%%%%%%%%%%%%%%%%%%%%%%%%%

%%%%%%%%%%%%%%%%%%%%%%%%%%%%%%%%%%%%%%%%%%%%%%%%%%
\section{Introduction}
\label{sec:introduction}

% There exist various choices for control barrier functions~(CBFs). Polynomial CBFs: synthesis through sum of squares~(SOS) programming for polynomial systems. For more general nonlinear systems, there often do not exist good synthesis solutions. One common choice for more general systems is a subset of the polynomial CBFs: affine or ellipsoidal functions. They offer a low complexity in terms of design. Because of their simplicity and commonality in practice, we explicitly consider ellipsoidal CBFs. 

% Another choice of CBF, is higher-order CBFs to handle the issue of control inputs not appearing in the Lie derivative. 
Safety filters have recently gained interest with the rise of learning-based control and reinforcement learning approaches.
While such learning-based approaches can improve the controller's performance based on interaction data, they do not provide safety guarantees (e.g., a self-driving vehicle that should stay inside lane boundaries). 
A safety filter tries to find the minimal modification to a potentially arbitrary control input proposed by an uncertified control policy that still achieves safety~\cite{DSL2021}. 
Safety filters typically rely on model knowledge of the system for accurate predictions~\cite{Fisac2019, zeilinger_linear, l4dc22}.
 
One popular safety filtering method relies on control barrier functions~(CBF). 
Control barrier functions can be used to encode control invariant sets. These sets guarantee that if the system's state is initialized in this set, there always exists a feasible control input to keep the system inside the set for all future time. 
If a control invariant set satisfies the state constraints, it is typically called a safe set, and the system is referred to as being safe inside this set~\cite{ames2019a}. 
The advantage of CBFs is that determining control invariance amounts to checking a scalar condition. 
For control-affine systems, this scalar condition can be used as an affine constraint in a quadratic program~(QP) to find the closest feasible control input to a proposed uncertified control input. This QP yields the control barrier function-based safety filter. 

\begin{figure}[t]
    \centering
    % This file was created with tikzplotlib v0.10.1.
\begin{tikzpicture}

\definecolor{darkgray176}{RGB}{176,176,176}
\definecolor{darkturquoise0191191}{RGB}{0,191,191}
\definecolor{gray}{RGB}{128,128,128}
\definecolor{green01270}{RGB}{0,127,0}
\definecolor{lightgray204}{RGB}{204,204,204}

\begin{axis}[
colorbar,
colorbar style={ylabel={$L_g h(x)$}, ytick={-80, -40, 0, 40, 80}},
colormap={mymap}{[1pt]
  rgb(0pt)=(0.403921568627451,0,0.12156862745098);
  rgb(1pt)=(0.698039215686274,0.0941176470588235,0.168627450980392);
  rgb(2pt)=(0.83921568627451,0.376470588235294,0.301960784313725);
  rgb(3pt)=(0.956862745098039,0.647058823529412,0.509803921568627);
  rgb(4pt)=(0.992156862745098,0.858823529411765,0.780392156862745);
  rgb(5pt)=(0.968627450980392,0.968627450980392,0.968627450980392);
  rgb(6pt)=(0.819607843137255,0.898039215686275,0.941176470588235);
  rgb(7pt)=(0.572549019607843,0.772549019607843,0.870588235294118);
  rgb(8pt)=(0.262745098039216,0.576470588235294,0.764705882352941);
  rgb(9pt)=(0.129411764705882,0.4,0.674509803921569);
  rgb(10pt)=(0.0196078431372549,0.188235294117647,0.380392156862745)
},
legend cell align={left},
legend style={fill opacity=0.8, draw opacity=1, text opacity=1, draw=lightgray204},
point meta max=86.8410282596298,
point meta min=-86.8410282596298,
height=4.7cm,
width=0.7\columnwidth,
tick align=outside,
tick pos=left,
x grid style={darkgray176},
xlabel={$x_1$},
xmin=-0.9, xmax=0.9,
xtick style={color=black},
y grid style={darkgray176},
ylabel={$x_2$},
ymin=-0.55, ymax=0.55,
ytick style={color=black}
]
\addplot graphics [includegraphics cmd=\pgfimage,xmin=-2, xmax=2, ymin=-2, ymax=2] {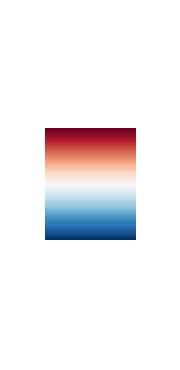};
\addplot graphics [includegraphics cmd=\pgfimage,xmin=-2, xmax=2, ymin=-2, ymax=2] {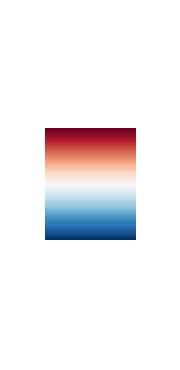};
% \draw[draw=gray,fill=gray,opacity=0.4] (axis cs:0,0) ellipse (0.874999999894803 and 0.49999999968437);
\draw[draw=black] (axis cs:0,0) ellipse (0.874999999894803 and 0.49999999968437);
% \addlegendentry{$h(x) = 0$}
\addplot [semithick, green01270, dashed]
table {%
-72.3675235496915 0
72.3675235496915 0
};
\node[] at (axis cs: 0.0, 0.05) {\textcolor{green01270}{\scriptsize $\text{Rel. deg.} > 1 \Rightarrow \text{Safety filter inactive}$}};
\node[] at (axis cs: 0.0, -0.44) {\scriptsize $h(x) = 0$};
\end{axis}

\end{tikzpicture}
    \caption{
    % Discrete-time implementation
    Visualization of the case study in \autoref{sec:simu-and-exp} for a linear system with an ellipsoidal CBF. 
    The states for which $\lVert L_gh(x) \rVert$ is small allow the CBF-based safety filter to apply control inputs close to the uncertified control input, rendering the safety filter close to being inactive. 
    Moreover, along the green line, the Lie derivative term $L_{g}h(x)$ is zero, and the relative degree $s$ does not equal one. In these states, the safety filter is completely inactive. 
    In a discrete-time implementation, the instantaneous inactivity of the safety filter can result in undesirable inputs that cannot be corrected at the subsequent discrete-time step.
    In both cases, this can lead to chattering and/or safe set violations. 
    % \todo{Illustration of a control barrier function~(CBF) that has different relative degrees throughout the certified region. If the relative degree of the CBF is misspecified, the} CBF-based safety filter can become inactive by entering certain parts of the state space. This leads to the application of uncertified control inputs to the system, which can eventually cause safe set violations for discrete-time implementations. 
    % % We highlight that this occurs for valid CBFs (see~\autoref{def:cbf}). 
    % The set of states where the safety filter becomes inactive are the states for which the Lie derivative of the input dynamics $g(x)$ with respect to the CBF $h(x)$ are zero \todo{TODO:has to use a higher order representation? $L_{g}L_{f}^{s-i} = 0$, where $i \in \set{N}$ and $i \leq s$ }: $L_g h(x) = 0$. Here, we show this set of states~(in green) for a two-dimensional system with a single input and an ellipsoidal CBF \todo{for relative degree $s = 1$}. Even though safety can be guaranteed when the safety filter is applied in a continuous-time setup, in discrete-time implementation, the instantaneous inactivity of the safety filter can result in undesirable inputs that cannot be corrected at the subsequent discrete-time step. When close to the safety boundary, this could lead to undesirable chattering effects and/or constraint violations.
    }
    \label{fig:money-figure}
\end{figure}
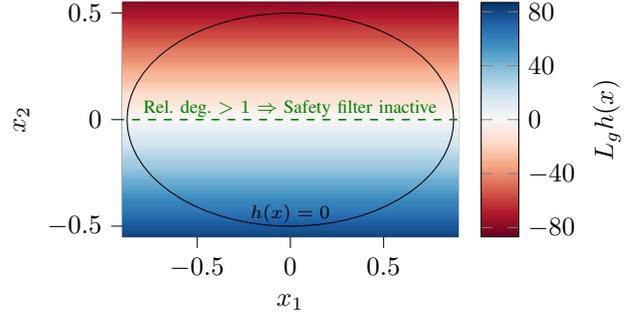

CBFs are typically defined for continuous-time systems. To preserve the CBF-based safety filter's guarantees, the aforementioned QP has to be solved infinitely fast~\cite{khojasteh2020a}. 
This is impossible in practice, such that the CBF-based safety filter only approximately provides safety guarantees by solving the QP at discrete timesteps. 
Nevertheless, applying this approximate strategy in practice has led to many successful real-world implementations, especially when the sampling time is small~\cite{gurriet-active-set-inv, Ohnishi2019, taylor2020a, singletary-food-2022, jian-dynamic-cbf-2023, brunke-lcss-2024}. 

A few works have systematically addressed this discrete-time implementation by analyzing the safety between discrete timesteps or providing an event-triggered formulation~\cite{khojasteh2020a, xiao-event-triggered-2023}. However, these approaches can lead to conservative closed-loop performance or \todo{rely on knowing the global relative degree. Checking the global relative degree is not always straightforward, and misspecification can lead to undesired behaviours, especially in discrete-time implementations. }
% providing either probabilistic guarantees that the safety filter certifies a control input that   paper on self-triggered implementation to determine the control input~\cite{khojasteh2020a}. Provides probabilistic guarantees on control invariance in between the time steps. 

There also exist discrete-time CBFs that have been specifically designed for discrete-time implementations~\cite{zeng2021safety, Cosner-discrete-cbf-2023, taylor2022-sampled}. However, even for control-affine systems, the resulting safety filter optimization problem is typically a nonlinear program, which reduces the real-time feasibility for real-world systems. 
% Continuous-time systems can be derived from physics. Discrete-time systems are approximations. 
Therefore, discrete-time implementations of continuous-time CBF-based safety filters continue to be a popular method for approximately guaranteeing the safety of real-world systems. 

Our contributions in this work are three-fold: \textit{(i)}
  we draw attention to practical issues of using standard continuous-time CBF-based safety filters in discrete-time implementations, \textit{(ii)} we propose three practical strategies to mitigate the issues, and \textit{(iii)} we verify our proposed strategies in simulation and real-world quadrotor experiments. 
%  \begin{itemize}
%     \item[(1)] We draw attention to two practical issues of using standard continuous-time CBF-based safety filters: 
% \textit{(i)} a misspecified global relative degree, and \textit{(ii)} the implementation in discrete time.
% \item[(2)] We propose three practical strategies to mitigate the issues.
% \item[(3)] We verify our proposed strategies in simulation and real-world quadrotor experiments. 
% \end{itemize}

% \todo{In this work, we draw attention to two practical issues of using standard continuous-time CBF-based safety filters: 
% \textit{(i)} a misspecified global relative degree, and \textit{(ii)} the implementation in discrete time. }
% After introducing the issues,
% % Our contributions in this work are three-fold: \textit{(i)} we draw attention to an issue for standard CBF-based safety filters implemented in discrete time that can cause safety violations, see~\autoref{fig:money-figure}, \textit{(ii)} 
% we propose three strategies to mitigate them and 
% % \textit{(iii)} 
% we verify our proposed strategies in simulation and real-world experiments. 

% \input{sections/related-literature}
\section{Problem Formulation}
\label{sec:formulation}
% We consider the control architecture in~\autoref{fig:blockdiagram}. 
% Our goal is to design a safety filter that provides safety guarantees to uncertified controllers in the presence of dynamics uncertainties and input constraints.
In this work, we consider the control architecture shown in Fig.~\ref{fig:blockdiagram} and a continuous-time nonlinear control system in the following control-affine form:
% The uncertain system has the following form:
\begin{equation}
\label{eq:nonlinear_affine_control}
	\xdot = \f(\x) + \g (\x) \:\u\, ,
\end{equation}
where $\x\in \set{X} \subset \R^n$ is the state of the system with $\set{X}$ being the set of admissible states, $\u \in \R^m$ is the input of the system, and  $\f:\R^n\mapsto \R^n$ and $\g:\R^n\mapsto \R^{n\times m}$ are locally Lipschitz continuous functions. We assume that $\set{X}$ is a known compact set, and $f$ and $g$ are known functions.

% \todo{formally define what safety means and then how constraint set is parametrized as a zero superlevel set}
% We define safety as 

% In this work, we assume that we have partial knowledge about the robot dynamics. Without loss generality, the dynamics of the robot system can be decomposed into a prior component and an unknown component as follows:
% \begin{equation}
%     \label{eq:disturbed_system}
%         \xdot = \fhat(\x) + \ghat (\x) \u + \vec{d}(\x, \u) \,,
% \end{equation}

% where $\vec{d}(\x, \u) = \vec{a}(\x) + \vec{b}(\x)^\intercal \u$ with $\vec{a} (\x) =  \f(\x) - \fhat(\x)$ and $\vec{b} (\x) = \g (\x) - \ghat (\x)$. We assume that the difference between the true and nominal system $\vec{d}(\x, \u)$ is bounded for all $(\vec{x},\vec{u})\in \set{X}\times \set{U}$.

% the system can be rendered positively control invariant on $\set{C}$. 
% We consider a given compact
The safe set $\set{C}\subseteq \set{X}$  is assumed to be given and is defined as the zero-superlevel set of a smooth function $h: \R^n \to \R$: $\set{C} = \{\x \in \set{X} \: \vert \: h(\x) \geq 0 \} $
% \begin{equation}
% % \label{eq:set_c}
%     \set{C} = \{\x \in \set{X} \: \vert \: h(\x) \geq 0 \} \,,
% \end{equation}
where the boundary of the safe set is $\partial\set{C} = \{\x \in \set{X} \: \vert \: h(\x) = 0 \}$ with $\partial h(\x) / \partial \x  \neq 0$ for all $\x \in \partial \set{C}$.
% , and the interior is $\text{Int}(\set{C}) = \{ x \in \set{X} \:\vert\: h(x) > 0\}$. 
% The corresponding unsafe set is defined as $\bar{\set{C}}=\{ x \in \set{X}\:\vert\: h(x) < 0\}$.
%
Our goal is to modify a given, potentially uncertified control policy~$\pi(x)$ with a safety filter based on a continuous-time CBF-based safety filter such that the system is safe (i.e., the system's state~$\x$ stays inside a safe set $\set{C}$ if it starts inside of $\set{C}$). 
%
% In this work, we particularly focus on the issues that arise when implementing a continuous-time CBF-based safety filter in practice, where the state measurements and input commands are only processed at discrete timesteps.
%
% \todo{discrete time implementation introduce here or later. our goal of this paper is to study one practical issue arose in implementing the CBF filter in practice, where the states, inputs are xx sampled intervals}
% \todo{still need to mention control invarinace somewhere, move the definition to the prelimnaries and say here h is a cbf?} BLANCHINI1999
% In this work, we assume a relative degree of $1$ for $h(\x)$.

% We assume that there exists a control input signal~$\nu: \mathbb{R}_{\ge 0}\mapsto \mathcal{U}$ such that the true system in~\autoref{eq:nonlinear_affine_control} stays inside $\set{C}$. 

% We also assume that there exists a control input signal~$\nu$, such that the nominal system in~\autoref{eq:nonlinear_affine_control} stays inside a subset $\set{C}_{\bar{h}}$ of the safe set $\set{C}$. 

\section{Background}
\label{sec:background}

\begin{figure}
    \centering
    \includegraphics[width=\columnwidth]{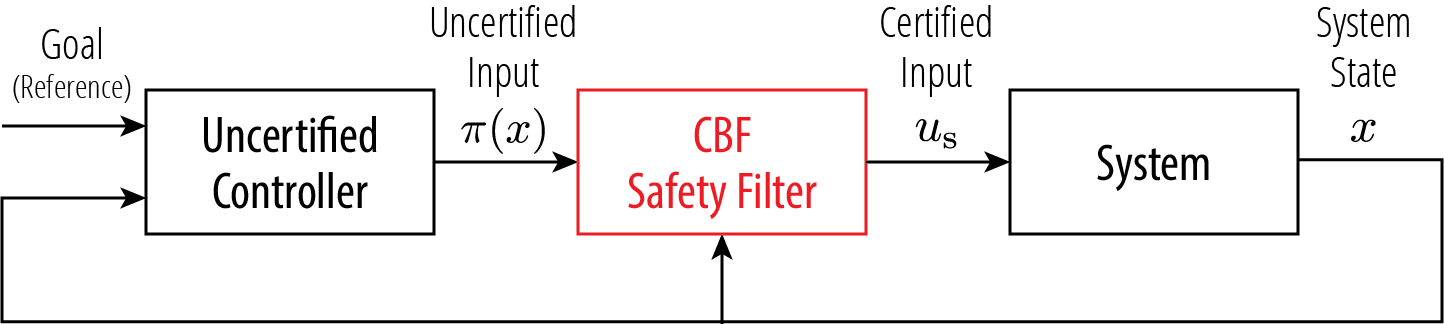}
    \vspace{0.1em}
    \caption{A block diagram of a typical CBF-based safety filter framework. A CBF-based safety filter is augmented to an uncertified controller~$\pi(x)$ and modifies the input of the uncertified controller if it is deemed unsafe.}
    \label{fig:blockdiagram}
\end{figure}

In this section, we introduce the necessary definitions and the relevant background on CBF-based safety filters~(see~\autoref{fig:blockdiagram}).
\begin{definition}[Extended class-$\K$ function~\cite{ames2019a}]
\label{def:kappa_inf_extended}
    A function $\gamma: \R \to \R$ is said to be of class-$\K_e$ %($\gamma \in \K_e$) 
    if it is continuous, $\gamma(0) = 0$, and strictly increasing. 
	% A continuous function $\gamma : \left(-b, a\right) \to \R$, with $\gamma(0) = 0$, $\gamma$ is strictly monotonically increasing, and $a, b = \infty$, $\lim_{r \to \infty} \gamma(r) = \infty$ and $\lim_{r \to - \infty} \gamma(r) = - \infty$, then $\gamma$ is said to belong to $extended~class~\K_\infty$ or equivalently expressed as $\gamma \in \K_{\infty, e}$. 
\end{definition}

\begin{definition}[Positively control invariant set] Let $\mathfrak{U}$ be the set of all bounded controllers $\nu : \R_{\geq 0} \to \R^m\,$. A set ${\set{C}\subseteq\set{X}}$ is a positively control invariant set for the control system in~\eqref{eq:nonlinear_affine_control} if~$\: \forall \: {\x_0 \in\set{C}} \,,\: \exists \: \nu \in \mathfrak{U} \,, \: \forall \: t \in \set{T}_{\x_0}^+ \,,\: \phi(t, \x_0, \nu) \in \set{C}$, where $\phi(t, \x_0, \nu)$ is the system's phase flow starting at $\x_0$ under the controller $\nu$, and $\set{T}_{\x_0}^+$ is the maximum time interval.
\end{definition}

\begin{definition}[Relative degree \cite{khalil2002}]
\label{def:rel-degree}
    The system consisting of the dynamics equation in \eqref{eq:nonlinear_affine_control} and the output equation $y = h(x)$ has relative degree $s \in \{1, \dots, n\}$ in $\set{X}_{s} \subseteq \R^n$ if it is $s$-th order differentiable and $L_g L_f^{i - 1} h(x) = 0$ for $i \in \{1, \dots, s - 1\}$ and $L_g L_f^{s - 1} h(x) \neq 0$ for all $x \in \set{X}_{s}$. 
\end{definition}

Intuitively, the relative degree specifies how often we have to differentiate $h$ along the dynamics \eqref{eq:nonlinear_affine_control} until the control input appears. If $\set{X}_s = \set{X}$, then $s$ is the global relative degree. Consequently, $s$ is a local relative degree if $\set{X}_s \subset \set{X}$. We use the global relative degree in the following two definitions for CBFs and higher-order CBFs. 

% The definition of a control barrier function following \cite{ames2019a} is given next. 
\begin{definition}[CBF~\cite{ames2019a}]
\label{def:cbf}
	Let $\set{C} \subseteq \set{X}$ be the superlevel set of a continuously differentiable function $h: \set{X} \to \R$, then $h$ is a CBF if there exists a class-$\K_e$ function $\gamma$ such that for all $\x \in \set{X}$ the control system in~\eqref{eq:nonlinear_affine_control} has a global relative degree of $s = 1$ and satisfies
	\begin{equation}
		\label{eq:cbf_lie_derivative}
		\max_{\u \in \R^m} \left[L_\vec{f} h(\x) + L_\vec{g} h(\x) \u \right] \geq - \gamma(h(x)) \,,
	\end{equation}
\end{definition}
where $L_f h(x) $ and $L_g h(x)$ are the Lie derivatives of $h$ along $f$ and $g$, respectively. In the following, we will also write $\dot{h}(x, u) = L_\vec{f} h(\x) + L_\vec{g} h(\x) \u$ for simplicity.
% \todo{introduce the CLF/CBF illustration (phase portrait) here?}
% \todo{
\begin{definition}[Higher-Order CBF\cite{xiao2021high}]
    \label{def:higher-order-cbf}
    Consider an $s$-th order continuously differentiable function $h: \set{X} \mapsto \set{R} $. Let $h_{i}(x), i \in \{1, ..., s\} $ be defined as 
    \begin{equation}
        h_{i}(x) = \dot{h}_{i-1}(x) + \gamma_{i}(h_{i}(x)), i \in \{1,...,s\}
    \end{equation} where $h_{0} = h$, and $\gamma_{i}(x), i \in \{1, ..., s\} $ denote class $\K_{e}$ functions
    and $\set{C}_{i}(t), i \in \{1, ..., s\}$ be defined by the superlevel set of $h_{i-1}, i \in \{1, ..., s\}$. Then $h$ is a Higher-Order CBF of relative degree $s$ if there exists differentiable class $\K_{e}$ functions $\gamma_{i}(x), i\in \{1, ..., s\}$ such that for all $x\in \set{X}$ the control system in~\eqref{eq:nonlinear_affine_control} has a global relative degree of $s$ and satisfies 
    \begin{align}
        &\max_{u\in \set{R}^{m}} \left[L_{f}^{s} h(x) +  L_{g}L_{f}^{s-1}h(x) u + O(h(x)) \right]\nonumber\\&\hspace{15em}\geq - \gamma_{s}(h_{s-1}(x)) %L_{f}^{s} h_{0}(x) + & L_{g}L_{f}^{s-1}h_{0}(x) u + O(h_{0}(x)) \right] \nonumber\\&\geq - \gamma_{s}(h_{s-1}(x)) 
    \end{align}
    for all $x \in \set{C}_{1}\, \cap, ..., \cap\, \set{C}_{s}$, where $O(h)$ is given by
    \begin{equation}
        O(h) = \sum_{i = 1}^{s-1} L_{f}^{i} (\gamma_{s-i} \circ h_{s-i-1} )(x).
    \end{equation}
\end{definition}
% \begin{equation}
    %     \set{C}_{i} = \{x\in \set{R}^{n}: h_{i-1}(x) \geq 0\}, i \in \{1, ..., m\}    
    % \end{equation}
    
Using a CBF \todo{with relative degree $s = 1$}, we can define an input set
\begin{equation}
    \label{eqn:cbf_input_set}
 \set{U}_{\text{cbf}}(x) = \{u\in\R^m \:\vert \:  \dot{h}(x,u) \geq - \gamma(h(x))\} \,
\end{equation}
that renders the system positively control invariant, which we will refer to as safe~\cite{ames2019a}.
This requires the selection of a class-$\K_e$ function~$\gamma$ that yields a non-empty $\set{U}_{\text{cbf}}(x)$ for all $x \in \set{C}$. For an uncertified controller $\pi(x)$ that may not designed to be safe, one can formulate a QP to augment the control input such that the system satisfies the CBF condition in~\eqref{eq:cbf_lie_derivative}~\cite{ Ames2014}:
\begin{subequations}
	\label{eqn:cbf_qp}
	\begin{align}
	\u_\text{s}(\x) = \underset{\u \in \R^m}{\text{argmin}} & \quad \frac{1}{2} \lVert \u - \pi(x) \rVert^2 \label{eqn:cbf_qp_cost} \\ \text{s.t.} & \quad \dot{h}(x, u) \geq - \gamma(h(\x))\,. \label{eqn:cbf_constraint}
	\end{align}
\end{subequations}
% An arbitrarily given feedback controller $\pi(x)$ might not guarantee that the system stays inside of the safe set $\set{C}$.
%satisfy the condition on the Lie derivative from~\eqref{eq:cbf_lie_derivative} of the control barrier function $h$. 
% The authors in \cite{freeman_kototovic, Ames2014} introduce a point-wise min-norm controller that minimally modifies the given controller to guarantee positive control invariance. 
% As the CBF condition on positive invariance 
% in~\eqref{eq:cbf_controller_set}
% is affine in the control input $\u$, we can define a quadratic program (QP) to find the minimum adjustment to a given controller $\pi(x)$ that renders the controlled system positively invariant:
% \begin{subequations}
% 	\label{eqn:cbf_qp}
% 	\begin{align}
% 	\u^*(\x) = \underset{\u \in \set{U}}{\text{argmin}} & \quad \frac{1}{2} \lVert \u - \pi(x) \rVert_2^2 \label{eqn:cbf_qp_cost} \\ \text{s.t.} & \quad \dot{h}(x, u) \geq - \gamma(h(\x))\,. \label{eqn:cbf_constraint}
% 	\end{align}
% \end{subequations}
%which uses the assumption $\set{U} = \R^m$. 
% Lipschitz continuity of the resulting controller $\u^*(\x)$ is shown in~\cite{Xu2015} \todo{check if this should rather be: B. J. Morris, M. J. Powell, and A. D. Ames. Continuity and smoothness properties of nonlinear
% optimization-based feedback controllers?}. 
%
% The application of the optimized controller $\u^*(\x)$ then satisfies the conditions on the Lie derivative of $h$ and the closed-loop system is also locally Lipschitz continuous. 
This is the safety filtering optimization that is solved inside the safety filter block in~\autoref{fig:blockdiagram}. Intuitively, the optimization problem in~\eqref{eqn:cbf_qp} finds an input in $\set{U}_\text{cbf}(x)$ that is as close as possible to $\pi(x)$, where the closeness of the inputs is specified with respect to a chosen distance measure~(e.g., the Euclidean norm in~\eqref{eqn:cbf_qp_cost}). \todo{For generic cases with a higher relative degree, we can define~\eqref{eqn:cbf_input_set} and~\eqref{eqn:cbf_qp} analogously.}
% In practice, the states of the system can only be measured or estimated at discrete times, and the above QP can only be solved in finite time $\Delta t > 0$. Consequently, the filtered input command is only 
% Therefore, a continuous-time implementation of~\eqref{eqn:cbf_qp} is infeasible in practice. 

In practice, a continuous-time implementation of~\eqref{eqn:cbf_qp} is often infeasible. This is partially due to the fact that the system state and input commands are only processed at discrete times. Moreover, practical systems are subject to delays (e.g., the time taken for solving the QP and communication delays in the control system), which hinders an ideal realization of a continuous-time CBF safety filter. In most implementations, the safety filter~\eqref{eqn:cbf_qp} is solved and applied to a system in a discrete-time fashion.

\section{Undesirable Effects of Discrete-Time Implementations of CBF-Based Safety Filters}
\label{sec:properties}

In this section, we discuss an undesirable effect of discrete-time implementations of continuous-time CBF-based safety filters. 

We consider the case, where~\eqref{eqn:cbf_qp} is solved at a sampling time~$\Delta t > 0$ to approximate the continuous-time safety filter policy. 
Consider the system~\eqref{eq:nonlinear_affine_control} at state $x(t_0)$ with $t_0 > 0$. Then, the certified control input is $u_{\text{s}}(x(t_0))$. The certified control input is applied over the time interval $t \in \left[ t_0, t_0 + \Delta t \right)$. However, only the initial time step $t_0$ has been certified by the safety filter, such that safety for the open time interval $t \in \left( t_0, t_0 + \Delta t \right)$ is typically not guaranteed in discrete-time implementations of continuous-time CBF-based safety filters. Poor performance in discrete-time implementations can especially arise when $ \lVert L_gL_f^{s - 1} h(x) \rVert \to 0$. In these cases, relatively large inputs are permissible by the CBF condition. %In discrete-time implementation, this could result in an undesirably large input applied over a sampling interval. 
However, when the system is close to the boundary, the large input applied over the finite time interval could result in subsequent chattering effects and even constraint violations.

%Besides the loss in guarantees, this type of implementation has led to many successful practical applications, especially when $\Delta t$ is chosen to be small~\cite{gurriet-active-set-inv, Ohnishi2019, taylor2020a, singletary-food-2022, jian-dynamic-cbf-2023}. 

Another related issue is when $L_gL_f^{s - 1} h(x) = 0$. This happens when the CBF has a local relative degree higher than~$s$, which can often be non-trivial to determine in practice. For a simple ellipsoidal CBF candidate $h(x) = 1 -  x^\intercal P x$ with $P \succ 0$, one would need to check $x^\intercal P g(x)$ is nonzero for all $x\in\set{X}$ to determine if the relative degree is one to satisfy the CBF definition. \autoref{fig:money-figure} shows an example where the relative degree at certain $x\in\set{X}$ is greater than one. At these states, the safety filter~\eqref{eqn:cbf_qp} becomes inactive (as $L_g h(x) = 0$). When $x \in \set{C}$ and $L_g h(x) = 0$, the lower bound on the Lie derivative~\eqref{eqn:cbf_constraint} is trivially satisfied for any control input $u \in \R^m$. Therefore, the control input can be chosen to minimize~\eqref{eqn:cbf_qp_cost}, such that $u_s(x) = \pi(x)$.  
While the Lie derivative $\dot{h}(x, u) = L_f h(x)$ is a constant for all $u \in \R^m$ at such a state $x$, we can have $g(x) \neq 0$ such that $\dot{x} = f(x) + g(x) u$ is nonzero for $u \in \R^m$. Although the Lie derivative is unaffected by the control input, the control input may still affect the closed-loop system dynamics. 
Therefore, at states $x$ where $L_g h(x) = 0$, the safety filter allows the application of the unsafe control input $\pi(x)$ for at least the time interval $\left[ t_0, t_0 + \Delta t \right)$. Since $\pi(x)$ may be any arbitrary control policy, this can lead to safe set violations, poor performance, or chattering (frequent switching between an active and inactive safety filter~\cite{Koller2019Learningbased, fpb-cdc2023}).

To avoid this undesirable closed-loop behaviour, we encourage practitioners to implement additional verification methods, either online or offline. We propose possible mitigation strategies in the next section.    

% lead to poor performance, yield chattering \todo{cite Fed}, and even result in a loss of safety when implemented in a discrete-time safety filter. 

% \begin{remark}
%     In~\cite{ames2019a}, higher-order CBFs are introduced such that $L_g L_f^{s - 1} h(x) \neq 0$ for all $x \in \set{C}$, where $s \in \N$ is the relative degree.
%     In this case, the problem discussed in this section does not occur using higher-order CBF conditions. 
%     A redefinition of CBFs with relative degree $s = 1$, such that $L_g h(x) \neq 0$ for all $x \in \set{C}$, would achieve the same result.  
%     However, this restricts the choice of higher-order CBFs as $L_f^s h(x)$ alone may already satisfy the higher-order CBF condition such that $L_g L_f^{s - 1} h(x) = 0$ can still lead to a valid higher-order CBF.   
%     If $L_g L_f^{s - 1} h(x) = 0$ is allowed for some states $x \in \set{C}$, then the problem discussed in this section can also occur for higher-order CBFs. Then, our proposed mitigation strategies in the next section also apply. 
% \end{remark}

\section{Mitigating Undesired Behaviors for Discrete-Time Implementations}
\label{sec:mitigation}

In this section, we discuss potential practical strategies for handling cases when  $ \lVert L_gL_f^{s - 1} h(x) \rVert \to 0$ and even $L_gL_f^{s - 1} h(x) = 0$~(which by \autoref{def:rel-degree} indicates a misspecified global relative degree). This list shall lend practitioners a set of methods to improve real-world implementations.  

One way to mitigate the issues resulting from $\lVert L_g L_f^{s - 1} h(x) \rVert \to 0$ is to let the sampling time~$\Delta t \to 0$. In the limit, this results in a continuous-time implementation. However, as discussed above, this is not feasible in practice.
Alternatively, a prediction model can be used to determine if a future time step is inside the safe set, e.g., $x(t + \Delta t) \in \set{C}$. However, this can result in nonlinear constraints in the safety filter optimization problem, e.g., using Euler integration and a nonlinear CBF. Therefore, we aim to provide mitigation strategies that minimally increase the online computation. 

\subsection{Penalty Term}
One method to handle the case of $\lVert L_gL_f^{s - 1} h(x) \rVert \to 0$ is by modifying the safety filter objective function by adding a term that explicitly accounts for $L_gL_f^{s - 1} h(x)$ becoming close to~0.
Our new proposed safety filtering objective is
\begin{equation}
\label{eq:new_objective}
    J(x) = \frac{1}{2} \lVert u - \pi(x) \rVert^2 + \frac{r}{2 \lVert L_gL_f^{s - 1} h(x) \rVert^2 } \lVert u - \pi_{\text{safe}} (x)\rVert^2 \,, 
\end{equation}
where $\pi_{\text{safe}}$ is a known safe backup control policy (e.g., a stabilizing controller that renders $\set{C}$ control invariant). %A possible choice is $\pi_{\text{safe}} = 0$, as this will turn the filtering optimization problem into a synthesis problem.  
The new objective~\eqref{eq:new_objective} replaces the standard CBF safety filtering objective for all $ \lVert L_gL_f^{s - 1} h(x) \rVert > \epsilon$ where $\epsilon$ is a small positive number and $r > 0$ is a weighting parameter. The closer $\lVert L_g L_f^{s - 1} h(x) \rVert$ gets to 0, the greater the impact of the second term in the safety filtering objective. In this case, the safety filter will track the safe backup control policy instead of the uncertified control policy~$\pi(x)$. The weighting parameter $r$ determines the balancing between the two terms when $ \lVert L_gL_f^{s - 1} h(x) \lVert$ is far from 0. To avoid numerical instabilities, we set $u_s(x) = \pi_{\text{safe}}(x)$ when we are in a state $x$ such that $ \lVert L_gL_f^{s - 1} h(x) \rVert \leq \epsilon$. 

This strategy requires almost no added computational effort. In practice, the design of the safe backup control policy will require some attention, such that the backup policy can return the system to states where $\lVert L_gL_f^{s - 1} h(x) \rVert > \epsilon$. Otherwise, the system will continue using the backup control policy~$\pi_{\text{safe}}$ for all future time. 

% In case $\frac{\partial h}{\partial x} g(x) \rVert = 0$, we set $u=0$. 
% Then, the closer the second part of the Lie derivative gets to $0$, the higher the weight of the control input being zero. \todo{Or to some other control input that tries to move the system away from this point? Maybe a safe control policy?  Otherwise, the system can get stuck if the normal to $PB$ (maybe define a set?) }

\subsection{Modified Safe Set Design}
The undesired behaviors can also be mitigated by accounting for the issue of $\lVert L_gL_f^{s - 1} h(x) \rVert \to 0$ during the design of the CBF for a system. In the following, we introduce two practical modification strategies. 

\subsubsection{Safe Set Transformation}
One method to avoid the effects of $\lVert L_gL_f^{s - 1} h(x) \rVert \to 0$ during the design is by applying a transformation to the CBF, e.g., a translation or a rotation, which yields $\Tilde{h}(x)$ and the new safe set $\Tilde{\set{C}}$, respectively. This transformation has to be chosen such that the closed-loop system from a specific initial condition $x_0$ is not affected by $ \lVert L_gL_f^{s - 1} \Tilde{h}(x) \rVert \to 0$ for the transformed CBF. 
For example, if the underlying control policy $\pi$ is known to be stabilizing for a certain subset of the safe set, we can safely use $\pi$ in those states, e.g., the inactive safety filter does not jeopardize safety. 
Therefore, the goal of this strategy is to find a transformation, such that the states $x$ for which $ \lVert L_gL_f^{s - 1} \Tilde{h}(x) \rVert \to 0$ are the states for which $\pi$ is safe.
The transformed CBF $\Tilde{h}(x)$ is given by
\begin{equation}
    \Tilde{h}(x) = h(R (x - \delta)) \,,
\end{equation}
where $\delta \in \R^n$ represents the translation and $R \in \text{SO}(n) \subset \R^{n \times n}$ represents the rotation matrix with the special orthogonal group $\text{SO}(n)$ such that $R$ satisfies $R R^\intercal = I$ with $I$ being the identity matrix and $\det R = 1$. Following the transformation, it is not guaranteed that $\Tilde{h}(x)$ is a valid CBF anymore and has to be verified before using it online \cite{brunke-lcss-2024}. 

% Using this approach, the set of states for which $\lVert L_gL_f^{s - 1} h(x)\rVert \to 0$ has been modified. 
Since it may be impractical to show that $\pi$ is stabilizing for states $x$ for which $\lVert L_gL_f^{s - 1} \Tilde{h}(x)\rVert \to 0$, there is also the option to  
% will exhibit undesirable behaviour. There are two possible ways to go about this in practice: Either the transformation shifts those states to states that will not be reached during online operation, or the transformation has to be combined 
combine the transformation with the penalty term as introduced in the previous subsection. 

% It may be difficult to predict the behaviour of the closed-loop system with the safety filter. Therefore, choosing the correct transformation may require some trials based on simulation.

\subsubsection{Safe Set Approximation}

Our final proposed method to handle $\lVert L_gL_f^{s - 1} h(x) \rVert \to 0$ is based on finding an alternative safe set~$\Tilde{\set{C}}$ that can be represented by a set of valid CBFs $\{ h_i(x) \}_{i = 0}^q$, where $q \in \N$ is the number of CBFs. The functions $h_i$ should be chosen in such a way that $ \lVert L_gL_f^{s - 1} h_i(x) \rVert > \epsilon$ for all $x \in \Tilde{\set{C}}$, where $\epsilon > 0$ is a parameter. 

One possible choice for such functions $h_i$ are affine functions of the form
\begin{equation}
    h_i(x) = p_i^\intercal x + b_i \,,
\end{equation}
where $p_i \in \R^n$ and $b_i \in \R$. This has the advantage that $\frac{\partial h_i}{\partial x}$ is independent of the state $x$ and constant. This can simplify the choice of $p_i$: pick $p_i$ such that $L_g h_i(x) = p_i^\intercal g(x) \neq 0$ for all $x \in \Tilde{\set{C}}$ such that a relative degree of 1 is achieved. However, 
%this may not always be feasible in practice. 
a disadvantage is that this results in a convex set $\Tilde{\set{C}}$, which can be restrictive. One may also let the alternative safe set be an inner approximation of the original safe set, i.e., $\Tilde{\set{C}} \subseteq \set{C}$ to satisfy the initial safety requirements. An improved approximation can be achieved by adding more constraints; the accuracy of the approximation and the computational demand have to be traded off in real-world applications. %. However, adding additional constraints also increases the computational effort. Therefore, the accuracy of the approximation and the computational demand have to be traded off in real-world applications. 

% \todo{Shall we define a set for when $\lVert L_g h(x) \rVert < \epsilon$?}

\section{Case Study: LTI System with Ellipsoidal CBF}
\label{sec:simu-and-exp}
In this section, we validate our proposed mitigation strategies in simulation and on a real-world quadrotor system. All of our CBF-based safety filters have been implemented using CasADi~\cite{Andersson2019}.

To demonstrate the undesirable effect, we first set up two examples~(in simulation and the real world) where the issues discussed in ~\autoref{sec:properties} occur. 
For both examples, we consider an ellipsoidal CBF of the form
\begin{equation}
    h(x) = \beta - (x - c)^\intercal P (x - c) \,,
\end{equation}
where $\beta > 0$ defines the superlevel set, $c \in \R^n$ is the ellipsoid's center, and $P \in \R^{n \times n}$ is a positive definite matrix. Here we select  $\beta = 1$, $P = \text{diag}(1.31, 4.00)$, $c = \begin{bmatrix} 0 & 0 \end{bmatrix}^{\intercal}$ for simulation and $c = \begin{bmatrix} 1.125 & 0 \end{bmatrix}^{\intercal}$ for real-world experiments.

\subsection{Simulation Results}
\label{sec:simulation}
The system of interest for the simulation is the following continuous-time LTI system
\begin{equation}
\label{eqn:system_example}
    \dot{x} = \underbrace{\begin{bmatrix}
        0.00 & 1.00 \\
        -0.09 & 0.10 
    \end{bmatrix}}_A x + \underbrace{\begin{bmatrix}
        0 \\ 18.09
    \end{bmatrix}}_B u\,, 
\end{equation}
which was identified from offline data collected on a real-world quadrotor system, with $x \in \R^2$ and $u \in \R$. 
This yields 
\begin{equation}
    L_g h(x) = - 2 x^\intercal P B \,.
\end{equation}
Note that, for all states $x \in \set{X}_{s \neq 1} = \{ x \in \R^2\: |\: x^\intercal P B = 0 \} $, we have $L_g h(x) = 0$ \todo{ with a relative degree $s \neq 1$; elsewhere we have $s = 1$}. 
For any state in the neighbourhood of $\set{X}_{s \neq 1}$, we have that $\lVert L_g h(x) \rVert$ is zero or close to zero. 
% In this example, all states $x \in \set{L} := \{ x \in \R^2\: |\: x^\intercal P B = 0, \forall \alpha \in \R \} $ which are the states $x \in \{ x \in \R^2\: |\: x =  \begin{bmatrix}
%      \alpha & 0 
% \end{bmatrix}^\intercal, \forall \alpha \in \R \} \neq \emptyset$. 
% Therefore, the issues discussed in section~\autoref{sec:properties} occur in our example as for an assumed global relative degree $s=1$, we misspecify the relative degree for states in $\set{L}$. 

% First, we show that the CBF $h$ is indeed a valid CBF. According to~\cite{ames2019a}, this requires $\frac{\partial h}{\partial x} \neq 0$ for any $x \in \partial \set{C}$. This is satisfied as $P$ is positive definite and $0 \notin \partial \set{C}$. 
% Furthermore, $L_g h(x)$ may not be 0 on the boundary of the safe set if $L_f h(x)$ is negative at those states. Otherwise, there will not exist a class-$\K_e$ function. In this example, $L_f h(x) = - 2 x^\intercal P A x = 0$ for all $x \in \set{L}$, and the overall Lie derivative is zero regardless of the input. For $x\notin \set{L}$, $L_gh$ is non-zero; for any $\gamma\in\K_e$, we can always find a $u$ that satisfies the CBF condition~\eqref{eq:cbf_lie_derivative}. Therefore, $h$ is a valid CBF for system~\eqref{eqn:system_example}. 

Here, we select the class-$\K_e$ function $\gamma$ as the identity map. The uncertified control policy is $\pi(x) = -0.1$. We simulate the system using a sampling time $\Delta t = \SI{0.001}{\second}$ for $\SI{15}{\second}$ and always initialize the system to $x_0 = \begin{bmatrix}
    0.5 & - 0.1
\end{bmatrix}^\intercal$. 

\begin{figure*}[tb]
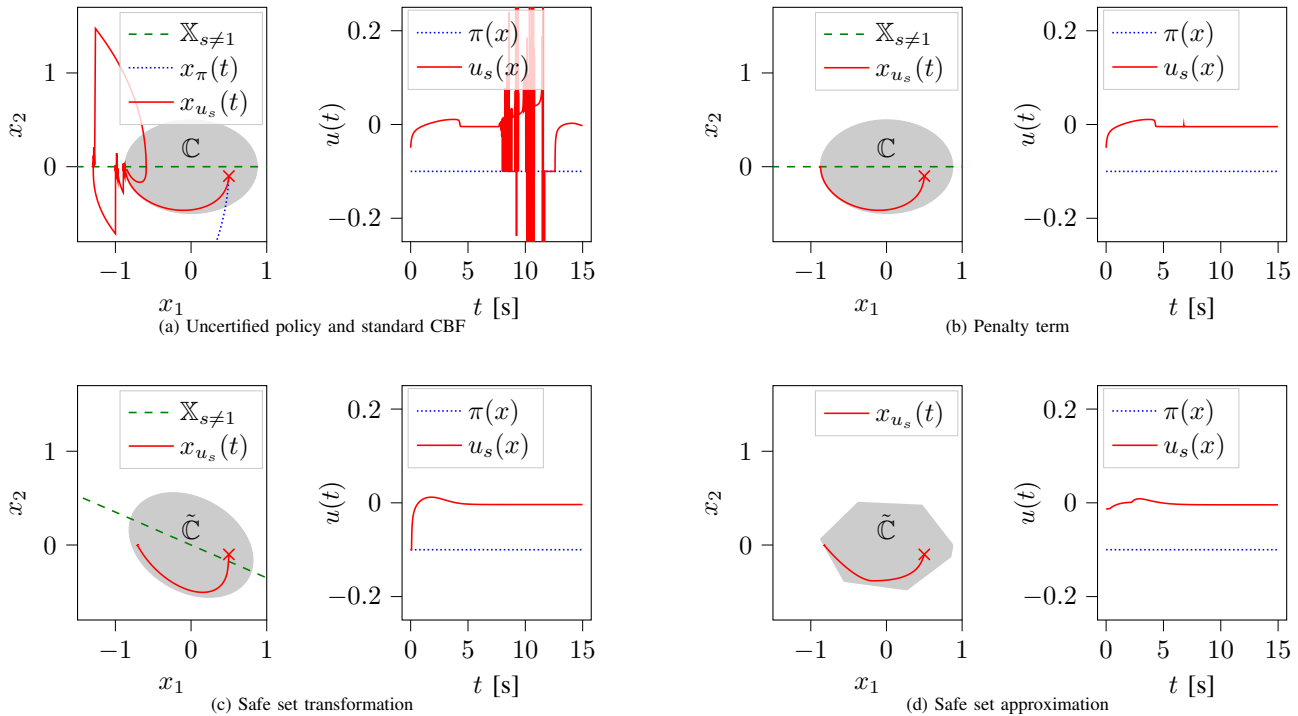

     \centering
     \begin{subfigure}[b]{0.48\textwidth}
         \centering
         \input{tikz/no-cbf}
         \vspace{-1em}
         \caption{Uncertified policy and standard CBF}
         \label{fig:uncertified-sim}
     \end{subfigure}
     \hfill
     \begin{subfigure}[b]{0.48\textwidth}
         \centering
         \input{tikz/penalty-cbf}
         \vspace{-1em}
         \caption{Penalty term}
         \label{fig:penalty-sim}
     \end{subfigure}\\
     % \hfill
     \vspace{1.3em}
     \begin{subfigure}[b]{0.48\textwidth}
         \centering
         \input{tikz/transformed-cbf}
         \vspace{-1em}
         \caption{Safe set transformation}
         \label{fig:transformed-sim}
     \end{subfigure}
     \hfill
     \begin{subfigure}[b]{0.48\textwidth}
         \centering
         \input{tikz/affine-cbf}
         \vspace{-1em}
         \caption{Safe set approximation}
         \label{fig:affine-sim}
     \end{subfigure}
     % \ref{mylegend}
        \caption{Demonstration of the undesired closed-loop behaviour when $\lVert L_g h(x) \rVert \to 0$ (see (a)) and our proposed mitigation strategies (see (b), (c), and (d)) in simulation. The closed-loop trajectories in (a, left) show the state trajectories using the uncertified control policy $\pi(x)$ and the certified control policy $\u_s(x)$, respectively. In both cases, the system leaves the safe set~$\set{C}$. For the certified control policy, this is caused by the closed-loop trajectory $x_{u_{s}}$ entering the neighbourhood of states where $L_g h(x) = 0$~(indicated by $\set{X}_{s \neq 1})$. This results in the system starting to chatter (see (a, right)) and eventually violating the safe set constraint~$\set{C}$. 
        In (b), we prevent safety violations by adding our proposed penalty term to the safety filtering objective. Initially, the closed-loop system behaves similarly to the standard CBF-based safety filter. Then, the system switches to a backup control policy $\pi_{\text{safe}} = 0$ when it enters a neighbourhood of $\set{X}_{s \neq 1}$.
        In (c), we successfully prevent chattering and safety violations by using a transformed safe set $\Tilde{\set{C}}$. This allows the closed-loop system to safely pass through the set $\set{X}_{s \neq 1}$ during the first couple of time steps and then stabilize at the final state in the simulation far from $\set{X}_{s \neq 1}$.  
        Finally, in (d), we demonstrate the mitigation of $\lVert L_g h(x) \rVert \to 0$ by using an alternative safe set~$\Tilde{\set{C}}$. Again, no safe set violations occur, as none of the affine constraints given by $h_i$ are parallel to the input matrix $B$. Therefore, $\set{X}_{s \neq 1}$ is empty.
        }
        \label{fig:sim-results}
\end{figure*}

The closed-loop state and input trajectories for the uncertified controller and the standard discrete-time implementation of a continuous-time CBF-based safety filter are shown in~\autoref{fig:uncertified-sim}. 
The system leaves the safe set~$\set{C}$ in both cases. For the certified control policy, this is caused by the closed-loop trajectory $x_{u_{s}}$ entering the set of states where ${L_g h(x) = 0}$~(indicated by $\set{X}_{s \neq 1})$. The CBF condition in~\eqref{eqn:cbf_constraint} certifies arbitrary control inputs for states $x \in \set{X}_{s \neq 1}$, rendering the safety filter inactive. This results in the system starting to chatter ($u_s$ ranging from $-3.20$ to $2.04$, see (a, right)) and eventually violating the safe set constraint~$\set{C}$.
% This shows that a valid CBF is insufficient for a CBF-based safety filter to keep the system inside the safe set. 
This is a result of the discrete-time implementation of the continuous-time safety filter.  

% \begin{figure}[tb]
%     \centering
%     \input{tikz/no-cbf}
%     \caption{Caption}
%     \label{fig:enter-label}
% \end{figure}

In the following, we apply our mitigation strategies to the simulation example. These strategies pursue one of two goals: \textit{(i)} prevent $ \lVert L_g h(x) \rVert \to 0$ by modifying the CBF, or \textit{(ii)} by using a safe control policy $\pi_{\text{safe}}$ when $\lVert L_g h(x)\rVert$ is close to 0. 
% Whenever the CBF is modified, we always verify that the modified CBF is still a CBF. 

First, we use the additional penalty term. Here, we select the new objective function to substitute~\eqref{eqn:cbf_qp_cost} with our new objective in~\eqref{eq:new_objective} 
% \begin{equation}
%     J(x) = \frac{1}{2} \lVert u - \pi(x) \rVert_2^2 + \frac{1}{2 \lVert L_g h(x) \rVert_2^2 } \lVert u \rVert_2^2 \,, 
% \end{equation}
and select $r = 1$ and $\epsilon = 10^{-8}$. The closed-loop state and input trajectories for the CBF-based safety filter with the modified objective function are shown in~\autoref{fig:penalty-sim}. 
Initially, the closed-loop system behaves similarly to the standard CBF safety filter. Unlike the standard CBF safety filter, our modified safety filtering objective results in a switch to the backup control policy $\pi_{\text{safe}} = 0$ when it enters a neighbourhood of $\set{X}_{s \neq 1}$. 
% This is a backup control policy, as all the states $x \in \set{X}_{s \neq 1}$ have $x_2 = 0$, so $x_1$ will be constant for all future time. 
Therefore, no chattering or safe set violations occur. 

The next mitigation strategy is achieved by transforming the safe set. For this example, we choose $\delta = 0$ and a two-dimensional rotation matrix~$R$ that rotates every state $x$ with an angle $\theta = \frac{\pi}{6}$ around an axis normal to the $x_1$-$x_2$ plane. \autoref{fig:transformed-sim} shows the closed-loop state and input trajectories for the CBF-based safety filter with the transformed CBF. 
The transformed CBF~$\Tilde{\set{C}}$ also modifies the set of states for which $L_g h(x) = 0$~(see dashed green line in~\autoref{fig:transformed-sim}). This allows the closed-loop system to safely pass through $\set{X}_{s \neq 1}$ and its neighbourhood while applying $\pi(x)$ during the first couple of time steps and then stabilize at a state $x(T)$ that is not close to $\set{X}_{s \neq 1}$, where $T > 0$ is the final time in the simulation. 
To achieve a useful transformation of the CBF, knowledge of which states yield $L_g h(x) = 0$ is required and necessitates an additional step in the offline design.
As mentioned above, this strategy can also be used with the additional penalty term. 

Finally, we find an alternative CBF to mitigate the undesirable impact of $L_g h(x) = 0$ on the closed-loop safety filtering behavior, see~\autoref{fig:affine-sim}. 
Again, no safe set violations occur, as none of the affine constraints given by $h_i$ are orthogonal to the set $B^{\perp} = \{ x \in \R^2 \,, x^\intercal B = 0 \}$. Therefore, this yields $L_g h_i(x) \neq 0 \,, i \in \{1, ..., 7 \}$ for all states $x$ in the modified safe set $\in \Tilde{\set{C}}$. We highlight that all the $h_i$ have to be specifically chosen to avoid $p_i^\intercal B = 0$ (for the affine case). 

The presented simulation results encourage determination for which states $L_gL_f^{s - 1} h(x) = 0$ for the problem at hand. This enables finding an effective modification to the safe set $\set{C}$. For high-dimensional systems or online adapted safe sets or systems (e.g., as in learning-based methods), the determination of $L_gL_f^{s - 1} h(x) = 0$ can be computationally infeasible. In such cases, the additional penalty term is a better option. 

% Then the partial derivative of the CBF~$h(x)$ with respect to $x$ is 
% \begin{equation}
%   \frac{\partial h}{\partial x} = - 2 (x - c)^\intercal P \,,
% \end{equation}
% and the system's Lie derivative is 
% \begin{equation}
%     \dot{h}(x, u) = - 2 (x - c)^\intercal P (f(x) + g(x) u)\,.
% \end{equation}

% In the following, we consider the null space of the second part of the Lie derivative~$\mathcal{N}_u = \mathcal{N}\left(\frac{\partial h}{\partial x} g(x)\right)$. Assuming that $h$ is a valid CBF according to (TODO: refer to Definition 3), then for any $x \in \mathcal{N}_u$, any choice of the control input~$u \in \R^n$ will satisfy the CBF condition. 
% For the definition of a CBF we have that $\frac{\partial h}{\partial x} \neq 0$. However, this still allows a nonempty~$\mathcal{N}_u$.

% A state $x$ is an element of the nullspace $\mathcal{N}_u$ if one of the following holds: \textit{(i)} $g(x) = 0$, \textit{(ii)} $x = c$, or \textit{(iii)} $P (x - c) \perp g_i(x)\,, \forall i \in \{1, \dots, m \}$, where $g_i(x)$ are the columns of $g(x)$. 

% Consider the case of $n = 2$ and $m = 1$. There exists a vector $x$ (always the case? This could also depend on $g(x)$. For constant $g(x)$ this is the case. ) that yields $P(x - c) \perp g(x)$. 

\subsection{Quadrotor Experiments}
We perform physical experiments on a miniature quadrotor, the Crazyflie 2.1~\cite{giernacki2017crazyflie}, to verify the proposed mitigation strategies. A video of the experiments can be seen at this link: \href{http://tiny.cc/practicalCBF}{\texttt{http://tiny.cc/practicalCBF}}. In the experiment, a CBF-based safety filter is supposed to prevent a falling quadrotor from colliding with the floor. We stabilize the attitude of the quadrotor such that the motion of the quadrotor is limited to its $z$-axis. Then the system's state reduces to $x = \begin{bmatrix} z_{\text{pos}} & z_{\text{vel}} \end{bmatrix}^{\intercal} \in \mathbb{R}^{2}$ and the applied input is the collective thrust $u_{\text{total}} \in \mathbb{R}$. The elements of the state vector, $z_{\text{pos}}$ and $z_{\text{vel}}$, are determined by a motion capture system and numerical differentiation, respectively. Instead of the collective thrust, we use the delta thrust as our control input $ u = u_{\text{total}} - u_{0} $ where $u_{0}$ is the collective thrust for hovering with mass $m = \SI{0.033}{\kilogram}$ and gravitational constant $g = \SI{9.81}{\frac{\meter}{\second^2}}$. This yields the following continuous-time LTI system:
\begin{equation}
\label{eqn:system_example_real}
    \dot{x} = \underbrace{\begin{bmatrix}
        0.00 & 1.00 \\
        0.00 & 0.00 
    \end{bmatrix}}_A x + \underbrace{\begin{bmatrix}
        0 \\ 30.30
    \end{bmatrix}}_B u 
\end{equation} 
% Analogously to the subsection on the simulation results, we can verify our CBF for the real-world experiment. 
The set $\set{X}_{s \neq 1}$ \todo{with relative degree $s \neq 1$} can be readily determined, and we note that the issue introduced in section~\autoref{sec:properties} also exists in our real-world system. The sampling time for the real-world experiments is~$\Delta t = \SI{0.167}{\second}$. We highlight that in our real-world system, the sampling time is much larger than in our simulation.
Recall that with a discrete-time implementation of a continuous-time CBF-based safety filter, safety is typically not guaranteed between consecutive sampling time instants. Therefore, more conservative, i.e., less steep, class-$\K_e$ functions have to be chosen to prevent the system from approaching the safe set boundary too fast. 

\begin{figure*}[tb]
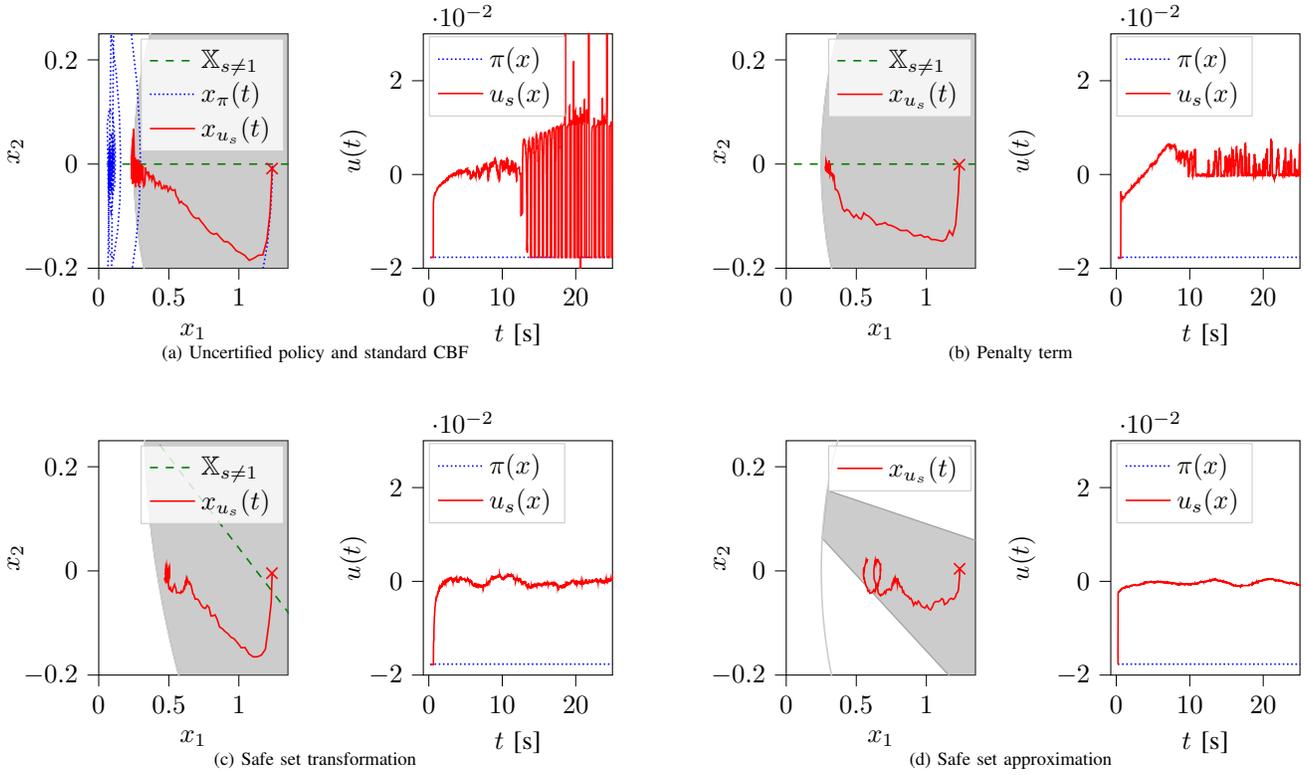

     \centering
     \begin{subfigure}[b]{0.48\textwidth}
         \centering
         \input{tikz/experiment-standard-cbf}
         \vspace{-1em}
         \caption{Uncertified policy and standard CBF}
         \label{fig:uncertified-exp}
     \end{subfigure}
     \hfill
     \begin{subfigure}[b]{0.48\textwidth}
         \centering
         \input{tikz/experiment-penalty-cbf}
         \vspace{-1em}
         \caption{Penalty term}
         \label{fig:penalty-exp}
     \end{subfigure}\\
     % \hfill
     \vspace{1.3em}
     \begin{subfigure}[b]{0.48\textwidth}
         \centering
         \input{tikz/experiment-transformed-cbf}
         \vspace{-1em}
         \caption{Safe set transformation}
         \label{fig:transformed-exp}
     \end{subfigure}
     \hfill
     \begin{subfigure}[b]{0.48\textwidth}
         \centering
         \input{tikz/experiment-affine-cbf}
         \vspace{-1em}
         \caption{Safe set approximation}
         \label{fig:affine-exp}
     \end{subfigure}
     % \ref{mylegend}
        \caption{Demonstration of the undesired closed-loop behavior when $\lVert L_g h(x)\rVert \to 0$ (see (a)) and our proposed mitigation strategies (see (b), (c), and (d)) on a real-world quadrotor system. The closed-loop trajectories in (a, left) show the state trajectories using the uncertified control policy $\pi(x)$ and the certified control policy $\u_s(x)$, respectively. The quadrotor violates the safe set~$\set{C}$ for both scenarios. For the standard CBF-based safety filter, chattering happens when the system enters the vicinity of set $\set{X}_{s \neq 1}$, which causes the quadrotor to leave the safe set.  
        In (b), with our proposed penalty formulation, the system switches to the backup control policy $\pi_{\text{safe}} = 0$ when it enters the neighbourhood of $\set{X}_{s \neq 1}$ and chattering is significantly reduced. Most importantly, the quadrotor strictly stays inside the safe set in this experiment.
        Then, in (c), when using a transformed safe set $\Tilde{\set{C}}$, no chattering and safety violations can be observed. The quadrotor ends up hovering inside of the transformed safe set $\Tilde{\set{C}}$. 
        In (d), we apply an alternative safe set~$\Tilde{\set{C}}$ to achieve $L_g h_i(x) \neq 0$ for all $x \in \Tilde{\set{C}}$. The quadrotor violates the new smaller safe set for a few states. However, the system stays inside the original safe set~$\set{C}$ throughout the experiment.
        % that still can be rendered safe by the other strategies. 
        No chattering is observed with this alternative safe set. 
        % Again, no safe set violations occur, as none of the affine constraints given by $h_i$ are orthogonal to the set $B^{\perp}$, which is orthogonal to the input matrix $B$.  
        }
        \label{fig:exp-results}
\end{figure*}

We begin our experiments by hovering at an approximated initial condition $x_{0} = \begin{bmatrix} 1.25 & 0 \end{bmatrix}^{\intercal}$. Then, the uncertified control policy $\pi(x) = -0.05mg$ is applied. In the different experiments on the real-world system, we demonstrate the closed-loop behavior of the system with no safety filter, the standard CBF-based safety filter, and the augmented CBF-based safety filters using each of our three proposed mitigation strategies. 

The experimental results of the closed-loop state and input trajectories for the uncertified controller and the standard CBF safety filter are presented in~\autoref{fig:uncertified-exp}. Both cases result in the quadrotor leaving the safe set. 
For the standard CBF safety filter, when the system enters the neighbourhood of set $\set{X}_{s \neq 1}$ where arbitrary control inputs are certified by~\eqref{eqn:cbf_constraint}, chattering can be observed with $u_s$ varying from $-0.14mg$ to $0.37mg$ at a high frequency. 
% The drone violates the safety set constraint $\mathbb{C}$. 

The proposed mitigation strategies are validated in the following experiments. We first add the penalty term to our objective function, which results in the new objective~\eqref{eq:new_objective}. In this experiment, we select $r = 75$ and $\epsilon = 10^{-8}$. When the system enters the neighbourhood of $\set{X}_{s \neq 1}$, a backup control policy $\pi_{\text{safe}} = 0$ is applied. 
This is a backup control policy, as all the states $x \in \set{X}_{s \neq 1}$ have $x_2 = 0$, so $x_1$ will be constant for all future time.
The state and input trajectories with this modified objective function are shown in~\autoref{fig:penalty-exp}. While chattering is not entirely prevented, its magnitude is significantly decreased with the certified control input $u_{s} \in \left[-0.0016mg, 0.02mg\right]$ during the timesteps when chattering occurs.
Furthermore, the safety constraints are not violated during the closed-loop operation. We note that for real-world safety-critical applications, the parameter $r$ can be picked conservatively so that the safety filter will tend to rather adopt the given safe backup control policy to avoid dangerous behaviors. 

Secondly, we apply the transformed safe set with the same design parameters as in the simulation subsection to reduce the unfavorable effect caused by the system's state entering the neighbourhood of $\set{X}_{s \neq 1}$. The real-world results are shown in~\autoref{fig:transformed-exp}. The quadrotor safely passes through the vicinity of set $\set{X}_{s \neq 1}$ and eventually, the augmented safety filter successfully achieves hover inside the transformed safe set $\Tilde{\mathbb{C}}$. Additionally, no chattering is observed because of the transformed safe set $\Tilde{\mathbb{C}}$. 

Finally, we leverage an alternative CBF in our last experiment. We select a new set of five affine constraints that satisfy $p_{i}^{\intercal}B \neq 0$ for $i = \{1, \dots, 5\}$ to accommodate the sampling effect of the discrete-time implementation on our real-world systems. The new safe set and system trajectories are shown In~\autoref{fig:affine-exp}. We observe no chattering but minor violations of the new safe set for a few states. After the violation happens, the quadrotor is brought back into the safe set by the safety filter and remains there for the rest of the experiment. The model mismatch between the real-world system and our LTI model or the large sampling time could cause this violation.
% due to the insufficient control frequency which again highlights the sampling effect. 
% In this case, a more conservative class $\K_{e}$ functions could be chosen to achieve safety for the new safe set. 
We also emphasize that, by choosing this alternative CBF, we sacrifice a large part of the original safe set, which is still rendered safe by this strategy. 
% So in this case, one may parallelly use the other strategies as a backup.

% \input{sections/experiments}
\section{Conclusion}
In this work, we highlight the issue of CBF-based safety filters becoming inactive in states where the norm of the Lie derivative of the CBF with respect to the input dynamics $\lVert L_gL_f^{s - 1} h(x) \rVert$ is close to zero~(or zero). 
In such states, the CBF-based safety filter certifies any control input as safe, including the uncertified control policy.
For discrete-time implementations, this inactivity can lead to chattering and/or safe set violations as an uncertified control input can be applied for the sampling duration. 

To prevent this issue, we propose three strategies to modify a standard CBF-based safety filter. 
Our first strategy aims at switching to a safe backup control policy close to states that cause the discussed issue using an additional term in the safety filtering objective.
The second strategy transforms the safe set such that the undesired states are avoided in closed-loop operation with the safety filter. 
Our last strategy leverages an alternative safe set that is carefully designed such that undesired states do not exist in the new safe set. % 
Finally, we demonstrate the presented issue and validate our three proposed mitigation strategies in simulation and on a real-world quadrotor. 

%%%%%%%%%%%%%%%%%%%%%%%%%%%%%%%%%%%%%%%%%%%%%%%%
\balance

\bibliographystyle{./IEEEtranBST/IEEEtran}
\bibliography{./IEEEtranBST/IEEEabrv,./references}

\balance
%%%%%%%%%%%%%%%%%%%%%%%%%%%%%%%%%%%%%%%%%%%%%%%%
\end{document}